\documentclass[graybox]{archivesofdatascience}
\usepackage[T1]{fontenc}
\usepackage{mathptmx}       
\usepackage{helvet}         
\usepackage{courier}        

\usepackage{graphicx}        
\usepackage[bottom]{footmisc}

\usepackage{amsmath}		
\usepackage{natbib}			

\usepackage{placeins}		
\usepackage{url}			
\usepackage{hyperref}		

\newcommand{\creativecommons}{\includegraphics[height=6pt]{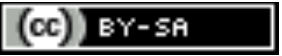}}

\begin{document}

\title*{Robust principal graphs for data approximation}
\author{Alexander N. Gorban$^1$\and Evgeny M. Mirkes$^1$\and Andrei Zinovyev$^2$ \\ \nobreakspace \\ \small{$^1$Department of Mathematics, University of Leicester, UK,\\ $^1$Institut Curie, France}}


\maketitle

\index{FIRSTAUTHORNAME, INITIAL@\emph{FIRSTAUTHORNAME, INITIAL}}
\index{SECONDAUTHORNAME, INITIAL@\emph{SECONDAUTHORNAME, INITIAL}}
\index{THIRDAUTHORNAME, INITIAL@\emph{THIRDAUTHORNAME, INITIAL}}
\index{data approximation}
\index{robustness to noise in data}
\index{principal graphs}
\index{manifold learning}
\index{elastic maps}
\index{trimmed approximation functional}

\abstract{Revealing hidden geometry and topology in noisy data sets is a challenging task. Elastic principal graph is a computationally efficient and flexible data approximator based on embedding a graph into the data space and minimizing the energy functional penalizing the deviation of graph nodes both from data points and from pluri-harmonic configuration (generalization of linearity). The structure of principal graph is learned from data by application of a topological grammar which in the simplest case leads to the construction of principal curves or trees. In order to more efficiently cope with noise and outliers, here we suggest using a trimmed data approximation term to increase the robustness of the method. The modification of the method that we suggest does not affect either computational efficiency or general convergence properties of the original elastic graph method. The trimmed elastic energy functional remains a Lyapunov function for the optimization algorithm. On several examples of complex data distributions we demonstrate how the robust principal graphs learn the global data structure and show the advantage of using the trimmed data approximation term for the construction of principal graphs and other popular data approximators.}

\section{Introduction}
\label{GORBANsec:1}
In this paper, we consider a classical problem: how to approximate a finite set $D$ in $R^m$ for relatively large $m$ by a finite subset of regular low-dimensional objects in $R^m$. In applications this problems arises  in many areas: from data visualization (e.g., visualization of the differences between human genomes) to fluid dynamics.


A typical data approximation task starts with the following question: whether the dataset $D$ is
situated near a low--dimensional affine manifold (plane) in $R^m$?
If we look for a point, straight line, plane, ... that minimizes the
average squared distance to the datapoints, we immediately come to
the Principal Component Analysis (PCA) which is one of the most
seminal inventions in data analysis \citep{jolliffe2002principal}.
The nonlinear generalization of PCA remains a challenging task.One of the earliest attempt
suggested in this direction was the Self-Organizing Maps (SOM) \citep{kohonen2001self} with its
multiple generalizations and implementations such as Growing SOM (GSOM) \citep{Alahakoon2000}.
However, unlike classical PCA and $k$-means, the SOM algorithm is not based on optimization of any explicit functional \citep{erwin1992self}.

For a known probability distribution, {\it principal manifolds} were
introduced as lines or surfaces passing through ``the middle'' of
the data distribution \citep{hastie1989}. Several  algorithms for
the construction of principal curves \citep{kegl2002} and surfaces for
finite datasets were developed during last decade, as well as many
applications of this idea. In the end of 1990s, a method of
multidimensional data approximation based on elastic energy
minimization was  proposed (see \citep{gorban1999neural,zinov2000,gorban2005elastic,Gorban2001ihespreprint,GorbanBook,GorbanHand} and the
bibliography there). This method is based on the analogy between the
principal manifold and the elastic membrane. Following
the metaphor of elasticity, two quadratic regularization terms are introduced
penalizing non-smoothness of data approximators.
This allows one to apply the standard expectation-minimization strategy
with quadratic form of the optimized functionals at the minimization step
(i.e., solving a system of linear algebraic equations with a sparse matrix).
Later on, the elastic energy was applied to constructing principal
elastic graphs \citep{gorban2007topological}. A related idea of optimizing the elastic energy of a system of springs representing
the graph embedment in low-dimensional spaces
was previously used in the development of graph drawing
algorithms (\cite{fruchterman1991graph,Kobourov2012}).

The method of elastic energy minimization allows creating analogs of SOM \citep{kohonen1982self} and neural gas \citep{martinetz1993neural} with an explicit functional to minimize: the elastic map is an analog of SOM and principal graph is an analog of neural gas. The main advantage of optimization-based analogs is the ability to explicitly control for smoothness (or other types of regularity, such as harmonicity) of data approximators.

However, the main drawback of all described methods of approximation is sensitivity to outliers and noise, which is caused by the very nature of Euclidean distance (or quadratic variance): data points distant to the approximator have very large relative contributions. There exist several widely used ideas for increasing an approximator's robustness in the presence of strong noise in data such as substituting the Euclidean norm by the $L_1$ norm (e.g. \citet{Ding2006, hauberg2014}) and outliers exclusion or fixed weighting or iterative reweighting during the construction of data approximators (e.g. \citet{Xu1995,allende2004robust,kohonen2001self}). In some works, it was suggested to utilize ``trimming'' averages, e.g. in the context of the $k$-means clustering or some generalizations of PCA \citet{cuesta1997,hauberg2014}).

The general idea of trimming consists in penalizing the contribution of distant from the mean value data points to the estimation of variance. In the simplest scenario the points that are too distant from the mean value are completely neglected; in more complex scenarios the distant points contribute less than the close ones. This way of robustification probably goes back to the notion of a truncated (or trimmed) mean value \citet{Huber1981}. The strategy of trimming can be used in the construction of SOMs, elastic maps or almost any other data approximators.


\section{Graph grammars and elastic principal graphs}
\label{GORBANsec:2}
Below in the description of basic algorithms we follow \citep{gorban2007topological}. More explanatory materials including the pseudo-codes can be found online\footnote{\url{https://github.com/auranic/Elastic-principal-graphs/wiki}}.

Let $G$ be a simple undirected graph with a set of vertices $Y$ and
set of edges $E$. For $k \geq 2$ a $k$-star in $G$ is a subgraph
with $k+1$ vertices $y_{0,1, \ldots k} \in Y$ and $k$ edges $\{(y_0,
y_i) \ | \ i=1,\ldots k\} \subset E$. Suppose for each $k\geq 2$, a
set of $S_k$ of $k$-stars in $G$ (subgraphs) has been selected. We call a graph
$G$ with selected families of $k$-stars $S_k$ an {\it elastic graph}
if, for all $E^{(i)} \in E $ and $S^{(j)}_k \in S_k$, the
correspondent elasticity moduli $\lambda_i > 0$ and $\mu_{kj}
> 0$ are defined. Let  $E^{(i)}(0),E^{(i)}(1)$ be vertices of an
edge $E^{(i)}$ and $S^{(j)}_k (0),\ldots S^{(j)}_k (k)$ be vertices
of a $k$-star subgraph $S^{(j)}_k $ (among them, $S^{(j)}_k (0)$ is the
central vertex).
 For any map $\phi:Y \to R^m$ the {\it energy of the
graph} is defined as
\begin{eqnarray}
U^{\phi}{(G)}&:=&\sum_{E^{(i)}} \lambda_i
\left\|\phi(E^{(i)}(0))-\phi(E^{(i)}(1)) \right\| ^2 \\ &&+
\sum_{S^{(j)}_k}\mu_{kj} \left\|\phi(S^{(j)}_k (0)) - \frac{1}{k}\sum _ {i=1}^k \phi(S^{(j)}_k
(i)) \right\|^2. \nonumber
\end{eqnarray}

For a given map $\phi: Y \to R^m$ we divide the dataset $D$ into
node neighborhoods $K^y, \, y\in Y$. The set $K^y$ contains the data points for
which the node $\phi(y)$ is the closest one in $\phi(y)$. The {\it
energy of approximation} is:
\begin{equation}\label{standardApproximationTerm}
U^{\phi}_A(G,D)= \frac{1}{\sum_{x} w(x)}\sum_{y \in Y} \sum_{ x \in K^y} w(x) \|x-
\phi(y)\|^2,
\end{equation}
where $w(x) \geq 0$ are the point weights. A simple and fast algorithm for minimization of the energy

\begin{equation}\label{globalStandardEnergy}
U^{\phi}=U^{\phi}_A(G,D)+U^{\phi}{(G)}
\end{equation}

\noindent is the splitting algorithm, in the spirit of the classical $k$-means clustering: for a given
system of sets $\{K^y \ | \ y \in Y \}$ we minimize $U^{\phi}$ (optimization step, it
is the minimization of a positive quadratic functional), then for a
given $\phi$ we find new $\{K^y\}$ (re-partitioning), and so on; stop when no change.

In practice the structure and complexity of the optimal graph for
approximation of a complex dataset is not known.
To learn it from the data themself, the principal elastic graphs
are constructed using a growing schema. All possible graph
structures are defined by a graph grammar. The optimal graph structure is obtained by the sequential application
of graph grammar operations to
the simplest initial graph \citep{gorban2007topological}. A link in the energetically optimal transformation chain is
added by finding a transformation application that gives the
largest energy descent (after an optimization step), then the next
link, and so on, until we achieve the desirable accuracy of
approximation, or the limit number of transformations (some other
termination criteria are also possible \citep{zinovyev2013data}).

As simple (but already rather powerful) example  we use a system
of two transformations: ``add a node to a node'' and ``bisect an
edge''. These transformations act on a class of {\it primitive
elastic graphs}:  all non-terminal nodes with $k$ edges are centers
of elastic k-stars, which form all the $k$-stars of the graph. This grammar
produces {\it elastic principal trees}, i.e. graphs having no loops.

\begin{figure}[tbp]
\includegraphics[width=0.9\textwidth]{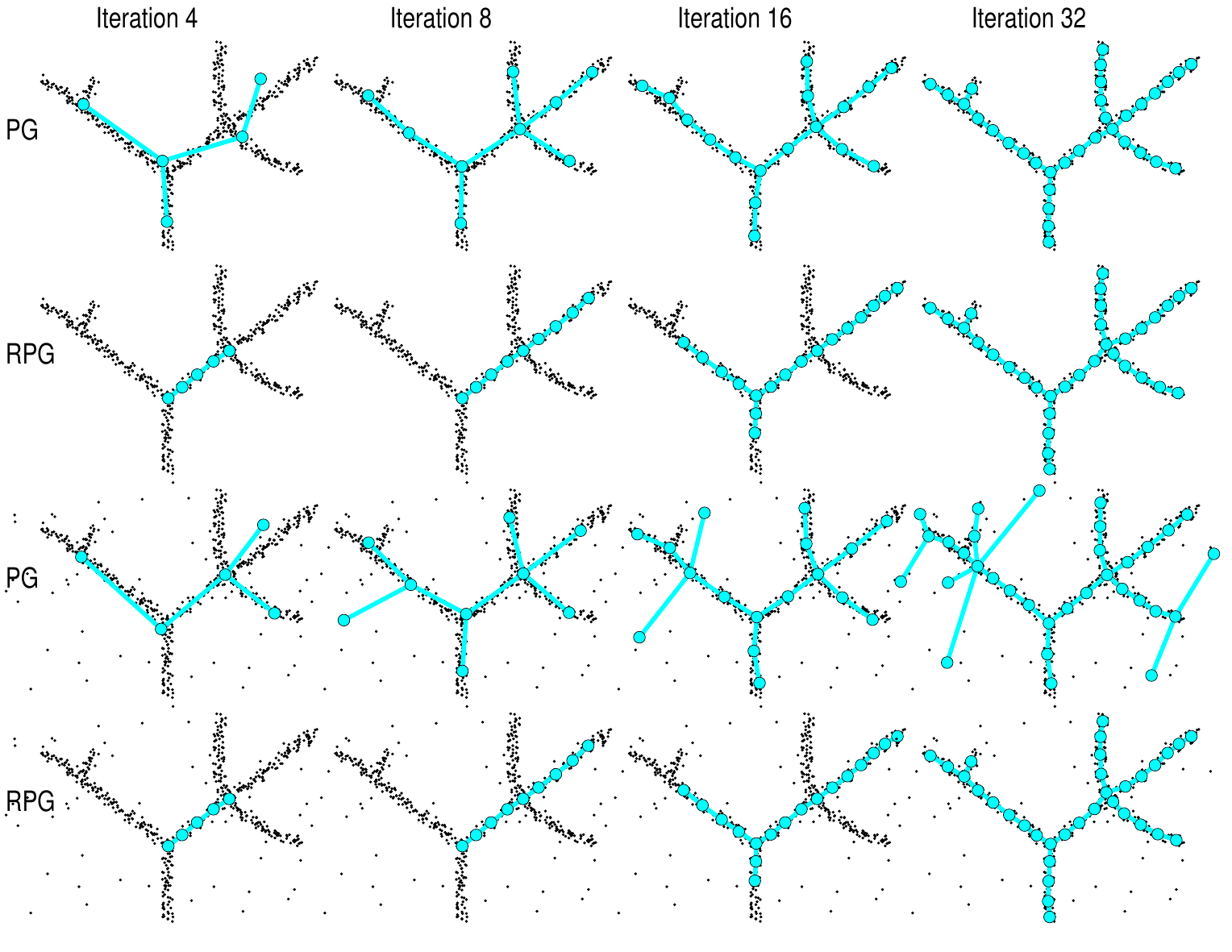}
\caption{Examples of principal graph (PG) and robust principal graph (RPG) construction for: clear (two top rows) and noised (two bottom rows) data patterns}
\label{GORBANEXAMPLE}       
\end{figure}

\section{Robust elastic principal graphs}\label{robustPGs}

In order to introduce the robust elastic principal graphs, we consider a motivating toy problem of learning a complex one-dimensional pattern sampled by
points densely located around. The pattern can be accompanied by a background noisy, relatively sparse and stochastic distribution of points not connected to the pattern
(see Figure~\ref{GORBANEXAMPLE}). The top row in Figure~\ref{GORBANEXAMPLE} shows several steps of the elastic principal graph construction in the case
of absence of noise.  From the Figure~\ref{GORBANEXAMPLE}, third row, it is clear that the presence of background noise can completely distort the
resulting principal graph, introducing excessive branching in order to capture the variance of the noise points located distantly from the pattern.

Here we apply a particular variant of ``impartial trimming'' \citep{gordaliza1991best} or ``data-driven trimming'' suitable for principal graphs.
We introduce a parameter $R_0$ (called ``robustness radius" further in the text) which specifies what is the maximal
distance from a node of the principal graph at which a data point can affect the position of the node during the current iteration of the energy optimization process.
We require that all the data points which are more distant than $R_0$ from any graph node, do not contribute into the gradient of the optimized functional $U^{\phi}$.
However, they have a constant non-zero contribution $R_0^2$ into the value of the data approximation term which is required for preserving the properties of the
optimized functional $U^{\phi}$ to be a Lyapunov function (see next section and Figure~\ref{FIGUREENERGY}).

In order to satisfy these requirements, we have to change the data approximation energy term only, because $U^{\phi}{(G)}$ term is independent of the data. The data-dependent approximation energy term (\ref{standardApproximationTerm}) is modified as following:

\begin{equation}\label{robustApproximationTerm}
U^{\phi}_R(G,D)= \frac{1}{\sum_{x} w(x)}\sum_{y \in Y} \sum_{x \in K^y} w(x) \min \{\|x-
\phi(y)\|^2,R_0^2\},
\end{equation}

\noindent where $R_0$ is the robustness radius. All other terms in the energy function are the same: $U^{\phi}=U^{\phi}_R(G,D)+U^{\phi}{(G)}$. 
It means that all optimization strategies used for construction of principal graphs are applicable for robust principal graphs too,
and that the optimization problem remains quadratic at the node optimization step. Notice that (\ref{robustApproximationTerm}) can be re-written as

\begin{eqnarray}\label{robustApproximationTerm2}
U^{\phi}_R(G,D)= \frac{1}{\sum_{x} w(x)}\sum_{y \in Y} ~~ \sum_{x \in K^y, ||x-\phi(y)||<R_0} w(x) \|x-
\phi(y)\|^2 + \\ + \frac{1}{\sum_{x} w(x)} \sum_{||x-\phi(y)||\geq R_0, \forall y} w(x)R_0^2, \nonumber
\end{eqnarray}

\noindent from which it becomes evident that the second term is constant and does not contribute to the derivative $U'_{y}$.


The result of such a modification is shown in Figure~\ref{GORBANEXAMPLE}, second and forth rows. Robust principal graph learns the data a local fragment and traces the local data structure, branching if this is energetically optimal. As a result, the global structure of the data distribution is detected only to the end of graph growth and only if there are no gaps in the data distribution larger than the robustness radius $R_0$.

\section{Convergence of robust elastic principal graphs}

Adding trimming in the data approximation term as in (\ref{robustApproximationTerm}) does not change the property of elastic principal graph's energy to converge to a local energy minimum. Both energy functions, the one defined by (\ref{globalStandardEnergy}) and the robust one

\begin{equation}\label{globalRobustEnergy}
U = \frac{1}{\sum_{x} w(x)}\sum_{y \in Y} \sum_{x \in K^y} w(x) \min \{\|x-
\phi(y)\|^2,R_0^2\}+U^{\phi}{(G)}
\end{equation}

\noindent are Lyapunov functions for the splitting-based optimization algorithm used for constructing the elastic principal graphs. The existence of Lyapunov function guarantees convergence of the optimization algorithm based on the splitting schema.

Let us formally demonstrate that at each step of optimization splitting algorithm, the energy (\ref{globalRobustEnergy}) does not increase.

Each graph optimization algorithm step is split into two parts. First, with fixed partitioning of the dataset $D$ into graph node neighbourhoods $K^y,y\in Y$, the quadratic function (\ref{globalRobustEnergy}) is minimized which leads to the new positions of nodes $\phi'(y)$. At this step, the energy $U$ can not increase because it is minimized: $U(\phi'(y))\leq U(\phi(y))$.

\begin{figure}[tbp]
\includegraphics[width=0.9\textwidth]{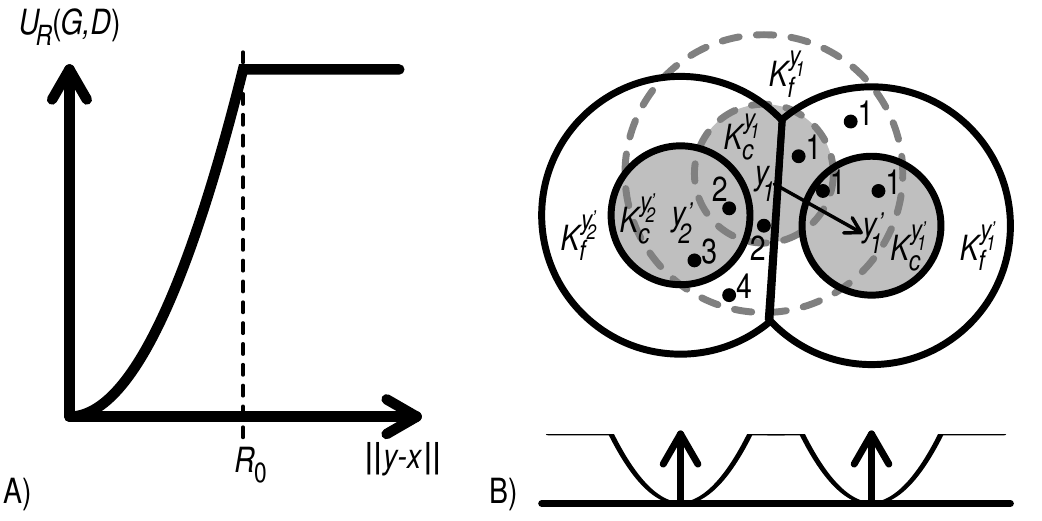}
\caption{A) The functional form of the trimmed approximation energy term. B) Possible re-partitioning of the data points (only several of them are shown as black circles with numbers) belonging to $K^{y_1}$ after changing the node position at the graph optimisation step. Four possible cases (1-4) are shown, and each one leads to non-increasing approximation energy (see text for explanation). Here the shaded area denotes the robustness radius $R_0$ for a graph node, the big solid circle denotes the neighbourhood $K^{y}$. The neighbourhood $K^{y}$ is split into ``close'' points $K_c^{y}$ and ``distant'' points $K_f^{y}$. }
\label{FIGUREENERGY}       
\end{figure}

Secondly, a new partitioning into sets $K^{y'}$ of data points $x$ is computed with respect to the new positions of nodes $\phi'(y)$. Let us denote by $K^y_c$ the set of points from $K^y$ which are not more distant than $R_0$ from $\phi(y)$: $K^y_c = \{x|\|\phi(y)-x\|\leq R_0\}$, and let us denote the set of ``distant'' points as $K^y_f = \{x|\|\phi(y)-x\|> R_0\}$. Of course, the whole neighbourhood is a union of these two sets, and their intersection is empty: $K^y = K^y_c\cup K^y_f$. After new partitioning, we will have a new $K^{y'} = K^{y'}_c\cup K^{y'}_f$. Let us consider one particular neighbourhood $K^{y_1}$ and any other one $K^{y_2}$. During the first step, $\phi(y_1)\rightarrow \phi'(y_1)$ and $\phi(y_2)\rightarrow \phi'(y_2)$. After re-partitioning, we might have several possible re-assignments of a data point $x\in K^{y_1}$  (see Figure~\ref{FIGUREENERGY}):

1. $K^{y_1}\rightarrow K^{y'_1}$: in this case the energy $U^{\phi}_R(G,D)$ does not change since the point $x$ remains in the neighbourhood of $y_1$.

2. $K_c^{y_1}\rightarrow K^{y'_2}$: in this case the energy term related to $y_2$ node in $U^{\phi}_R(G,D)$ decreases because by the definition of a neighbourhood $\|\phi'(y_1)-x\|>\|\phi'(y_2)-x\|$.

3. $K_f^{y_1}\rightarrow K_c^{y'_2}$: in this case the energy term related to $y_2$ node in $U^{\phi}_R(G,D)$ will not increase because it will change from $R_0^2$ to $\|\phi'(y_2)-x\|^2 \leq R_0^2$.

4. $K_f^{y_1}\rightarrow K_f^{y'_2}$: in this case the energy term related to $y_2$ node in $U^{\phi}_R(G,D)$  does not change (it equals $R_0^2$ before and after re-partitioning).

The same four scenarios are valid for any pair $y_i\neq y_j$: therefore, the total energy $U^{\phi}_R(G,D)$ can not increase while $U^{\phi}{(G)}$ is not affected by re-partitioning. It is also clear that the non-negativity of the derivative of the trimmed approximation energy function (Figure~\ref{FIGUREENERGY}A) is essential for that $U$ does not increase, because otherwise case (3) can lead to increase of $U^{\phi}_R(G,D)$. For example, disregarding ``distant'' points completely for their contribution to the approximation energy would lead to a violating property (3).

\section{Comparing various types of robust and non-robust data approximators}\label{Result}

\begin{figure}[tbp]
\includegraphics[width=0.9\textwidth]{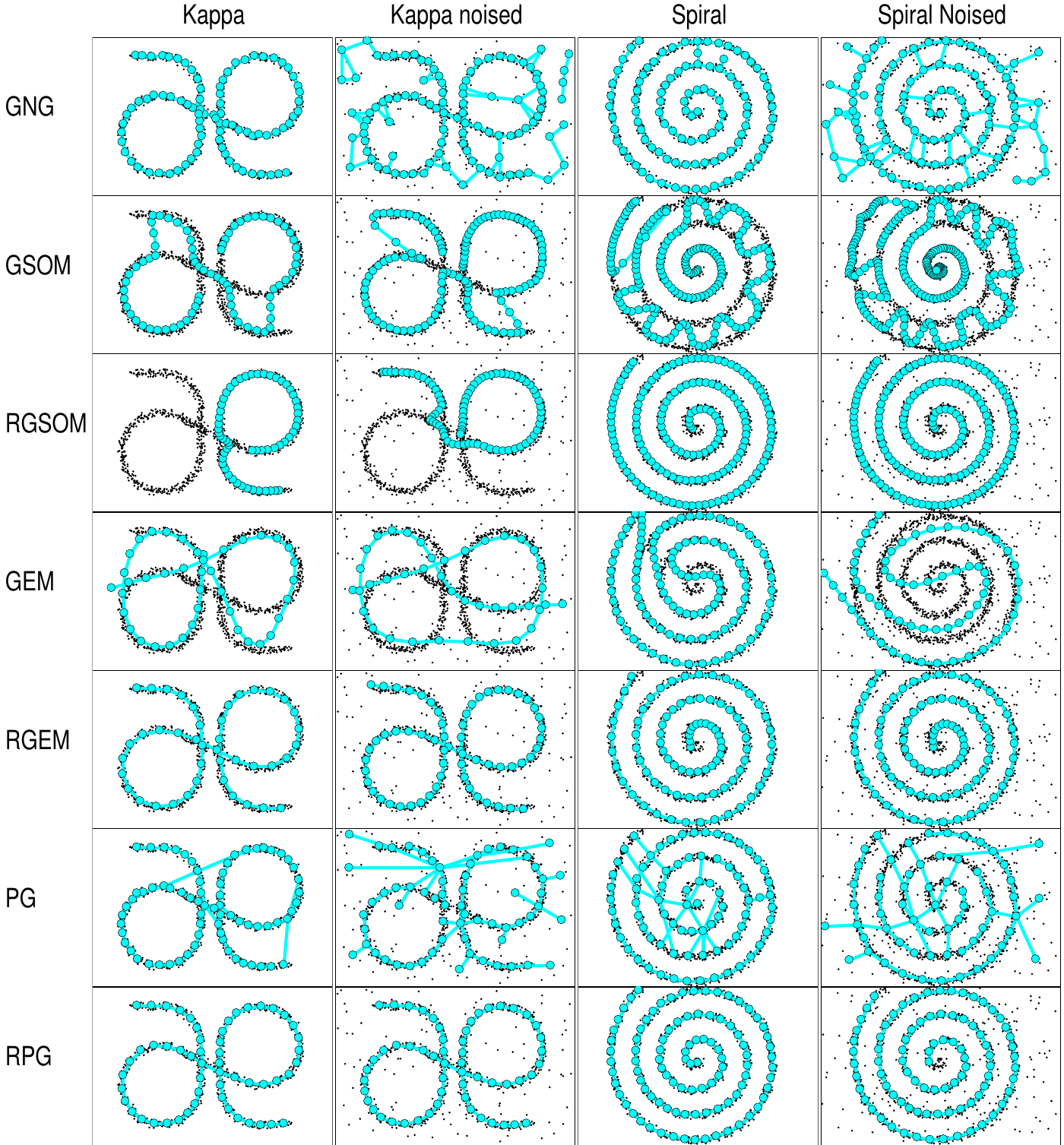}
\caption{Examples of different data approximator performances for four 2D toy complex patterns: from left to right: clear kappa-like, noised kappa-like, clear spiral, noised spiral; from top to bottom: Growing Neural Gas (GNG)\citep{martinetz1993neural}, Growing SOM (GSOM) and Robust GSOM (RGSOM) \citep{Alahakoon2000}, growing elastic map (GEM), robust GEM (RGEM), principal graph (PG)\citep{gorban2007topological,GorbanHand}, and robust principal graph (RPG). The parameters of computational protocols are provided in section \ref{Implementation}}.
\label{GORBANROTHER}       
\end{figure}

Let us further exploit the benefits that trimming the data approximation term can bring to approximating complex toy 2D patterns. Figure~\ref{GORBANROTHER} shows the results of application of standard and the robust versions of SOM, elastic maps and principal graphs for spiral and kappa-like data 2D patterns. 10\% of noise is introduced into the data distribution, e.g. the fraction of randomly positioned points not belonging to the data pattern is 0.1 of the number of points in the pattern. For robustification of the non-batch SOM, some specific methods were used described in \citet{allende2004robust} and \citet{kohonen2001self}. All robust versions of methods use the same robustness radius. In cases of all methods, robustification of approximators led to more exact approximation of the data pattern.

\section{Application of robust graphs to the data on human genome diversity}

Human Genome Diversity Project (HGDP) collected a large collection of single nucleotide polymorphism (SNP) genotype profiles capturing the major sources of variation between human genomes. The publicly available HGDP dataset which can be downloaded from \url{http://www.hagsc.org/hgdp/files.html} contains samples of 53 historically native population from 7 large geographical regions (Africa, Near East, Europe, South Central Asia, East Asia, America, Oceania). Each of 1043 individuals in this dataset is represented by a profile of 660918 SNPs. It was demonstrated before that the dataset shows non-trivial branching structure reflecting combined effect of migration and adaptation of humanity in various geographical conditions\citep{Elhaik2014}. Therefore, it is interesting to approximate this dataset by an optimally branching approximator such as the principal tree.

In order to represent the dataset as a set of numerical vectors in a multidimensional space, we've applied the standard SNP quantification approach. For each row of the table corresponding to a particular SNP, the homozygous status of the SNP was assigned '0' value, while two different heterozygous statuses were assigned '-1' and '+1' values. All unreliably measured SNP statuses were filtered out, which resulted in a numerical table of 1043 individuals (objects) and 429830 SNPs (variables). At first, we've reduced the dimension of the dataset to $R^3$ by applying the standard principal component analysis (PCA). In this reduced space we've constructed both standard and robust versions of principal trees.

Application of the standard principal trees was not successfull due to inability to capture fine local details of the data distribution (the figures can be found online\footnote{\url{http://goo.gl/CbFMlC}}), which resulted in mixing up the native populations belonging to the same region. For example, all variants of European population were mapped in a single tree node. Changing the elasticity parameters did not improve the situation.

At the same time, application of the robust principal trees with initialization from two points belonging to a local neighborhood resulted only to local description of the variety of human genomes. This was probably due to the fact that the distribution of individuals is characterized by certain gaps not covered by 1043 genomes (and which could probably represent non-existent part of human genome diversity).

\begin{figure}[tbp]
\includegraphics[width=1\textwidth]{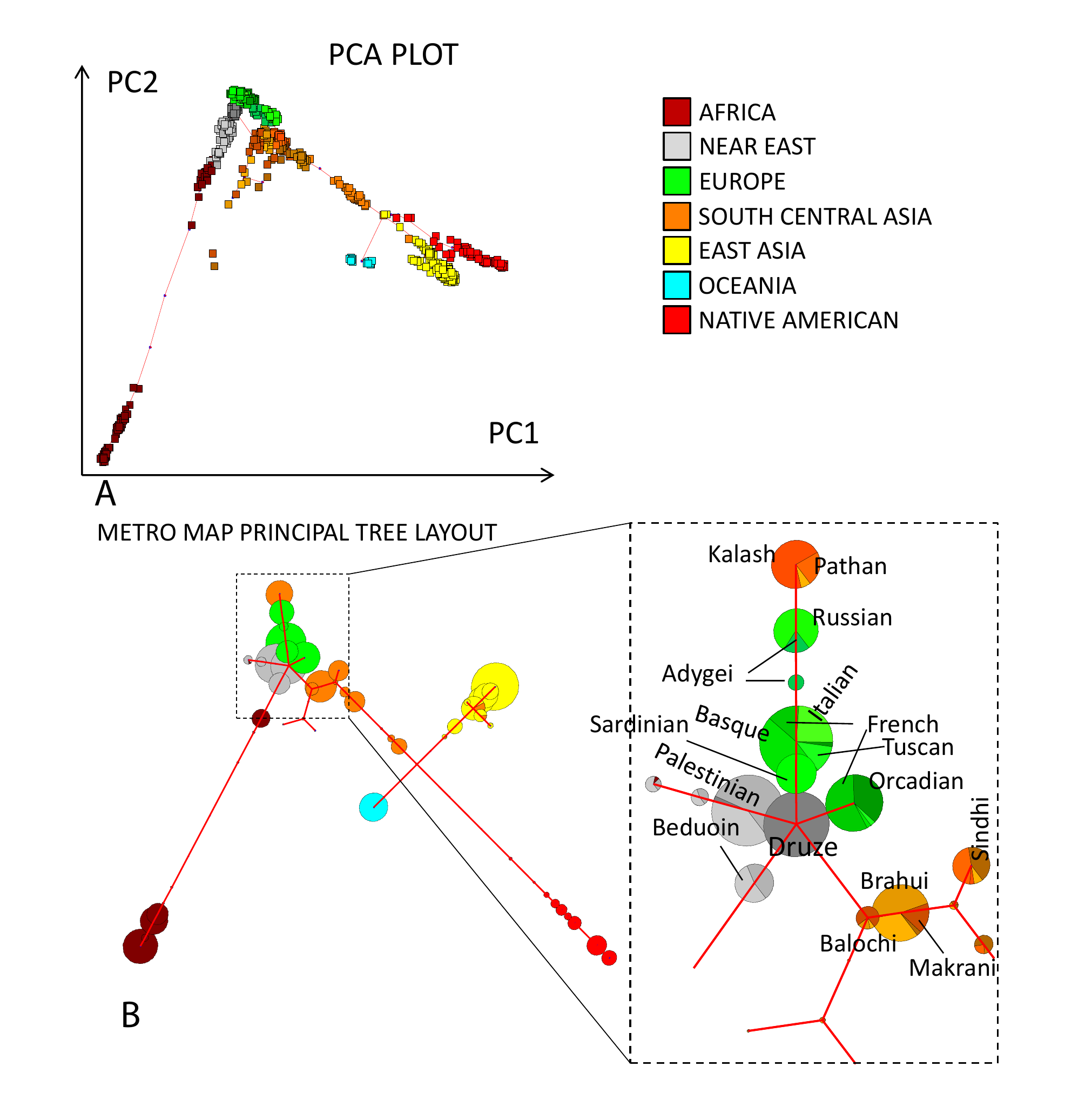}
\caption{Constructing a robust principal tree for the dataset mapping human genome diversity measured by single nucleotide polymorphism (SNP) genome-wide profiles (HGDP dataset). Large geographical regions are presented by a color while various tints of the color represent the distinct native populations within a region (only variability of European, Near East and part of South Central Asia is shown by aspects of the same color). A) Principal component analysis of the HGDP data, mapping data points from $R^{429830}$ to $R^{2}$. Red line shows the embedding of the constructed robust principal tree into the data space. B) Robust principal tree is shown using 2D metro map layout (planar harmonic representation of a tree). A particular region of the graph is shown zoomed on the left.}
\label{GeoSNPExample}       
\end{figure}

Therefore, it was not possible to find a combination of parameters which would represent both global and local patterns of the branching distribution of human genomes distribution. 
However, we could significantly improve the result by application of a hybrid approach. 
We trained the principal tree in two epochs. During the first epoch,  non-robust principal graphs were applied to outline the general features of the global structure of the dataset, representing roughly the global relations between geographical regions. 
By contrast, during the second epoch, robust principal tree approach was applied starting from the principal tree configuration obtained at the first step. 
During the second stage, the elasticity coefficients of the principal tree were significantly reduced in order to achieve better local approximation of the data. 
As a result, the constructed principal tree was able to capture the global pattern of genomic diversity between geographical regions and the local patterns of genomic diversity between native sub-populations of the same region. 
For example, Russian, Italian, Sardinian, Orcadian, Druze, Basque, Kalash populations were mapped to their own nodes (see Figure~\ref{GeoSNPExample}), resolving the structure of genomic diversity at a higher level of details.

In order to represent the principal tree on a 2D plane we used the previously described metro map layout of a tree on the plane\citep{gorban2008beyond,GorbanPMaGiP}. This layout is constructed in order to represent in the best fashion the harmonical nature of the embedding of a principal tree into multidimensional space. 
The center of each star of the tree is the mean point of the set of the star's leaves (see Figure~\ref{GeoSNPExample}B). 
The number of individuals from distinct populations mapped to the same tree node were represented as a pie-chart diagram(Figure~\ref{GeoSNPExample}B).
The distances between tree nodes in 2D represent approximately the edge lengths in the multidimensional space, 
therefore, it is possible to estimate which populations are more distinct than others. 
For example, unlike 2D PCA plot (Figure~\ref{GeoSNPExample}A), from metro map layout it is clear that the composition of SNPs of native Americans is significantly different 
from the one of Asian populations, even more distinct that the SNP profiles of population of Oceania. 
Interestingly, a part of East Asia population (Kalashs and Pathans) was mapped closer to the European populations than the rest of East Asia. 
Indeed, there exists a controversial discussion of whether Kalash and Pathan populations living in Pakistan have European roots 
(e.g., originated from the troops of Alexander the Great)\citep{Firasat2007,wood2001footsteps}. 
Here we can not make any strong conclusion regarding this point: however, construction of principal trees could potentially contribute to similar discussions.

To conclude, we can observe that introducing robust principal trees in the analysis of genomic data allows better tracing the local patterns in the complex real-life data distributions.

\section{Implementation details and computational protocols}\label{Implementation}

All 2D illustrations used in this paper are created by a Java applet, developed by the authors, for constructing non-linear approximations of 2D data, using various algorithms \citep{Applet}.  The construction of principal graphs in multidimensional space was performed using VDAOEngine Java library developed by the authors. The Parameters of the methods used are provided together with the code from the corresponding GitHub page\footnote{\url{https://github.com/auranic/Elastic-principal-graphs/wiki/Robust-principal-graphs}}.

\section{Conclusion and Summary}\label{Discussion}

We propose a robust version of a principal elastic graph algorithm which is based on trimming the data approximation term in the elastic energy of the graph. 
Growing principal graphs proceeds by approximation of local data structures and tracing them till the global structure is detected. 
For those data distributions which contain several isolated clusters, it is necessary to restart robust principal graphs several times
(one graph for each cluster) or apply a hybrid approach described in this paper. The algorithm contains an additional parameter $R_0$ which is called the robustness radius. 
Only the data points inside this radius around a graph node $y$ can influence position of $y$ at the next iteration step. 
The algorithm shows good performance in the case when the global data structure is spoiled by noisy background distribution of data points, 
which makes the algorithm more suitable in many practical applications (such as image recognition). The existence of Lyapunov function guarantees convergence 
of the optimization algorithm based on the splitting schema.

The suggested data trimming can be applied for other data approximators such as elastic maps and SOMs. 
In the future, we plan to apply the recently suggested machine learning framework (PQSQ-based optimization) to introduce a piece-wise quadratic 
form of the data approximation term which will allow taking into account the position of distant data points with smaller weights and avoid the problem
of defining the robustness radius \citep{Gorban2016NeuralNetworks}.

\FloatBarrier
\bibliographystyle{spbasic}
{
\FloatBarrier
\bibliography{GMZECDA2015}
}

\end{document}